# UAV-based detection of landmines using infrared thermography

**Muhammad Umair Akram Butt** *
Department of Electrical Engineering,
National University of Computer and Emerging Sciences, NUCES-FAST, Pakistan
Email: umair.akram@nu.edu.pk

**Zaighum Naveed**
Institute of Mechanical and Electrical Engineering,
University of Southern Denmark, Sønderborg, Denmark

**Usama Javed**
Department of Electrical Engineering,
National University of Computer and Emerging Sciences, NUCES-FAST, Pakistan

## Abstract

Landmines remain a pervasive threat in conflict-affected regions worldwide, exacting a toll on innocent lives. Shockingly, every 95 minutes, another individual becomes a victim of these lethal explosive devices (Landmines Monitor 2022 2022), with a significant proportion being innocent civilians. Current methods for landmine detection suffer from inefficiency, high costs, and risks to the operator and system integrity. In this paper, we present a novel, efficient, safe, and cost-effective approach to unearth these hidden dangers. Our proposed method integrates an unmanned aerial vehicle (UAV) with a thermal camera to capture high-resolution images of minefields. These images are subsequently transmitted to a base computer, where a state-of-the-art image processing algorithm is applied to identify the presence of landmines. Notably, our solution performs exceptionally well, particularly during evening hours, achieving an impressive detection accuracy of nearly 88%. These results exhibit great promise when compared to existing methods constrained by their design limitations.

## 1. Introduction

Landmines are explosive devices carefully packed inside a container. They are buried on or a few centimetres beneath the ground. The primary aim of this device is to kill or damage enemy personnel or vehicles. Due to their cheapness, effectiveness, and self-detonating ability, the landmines are abundantly used around the globe. Currently, 80 million landmines are garrisoned in many countries, and more than double, that is, 180 million are in the stockpile (Landmine Facts 2016). The cost to remove all the landmines is expected to be around $50 to $100 billion (Facts About Landmines 2018).

*Corresponding author

Thus, demining has been a very challenging task for many countries around the globe. Many demining techniques have been developed so far. Conventional methods include manual detection, metal detectors (F.Y.C. Albert 2014), the use of biological species (DeAngelo 2018); (Adee Schoon 2022); (Shubha Rani Sharma 2019); (Manley 2016); (Benjamin Shemer 2021); (Janja Filipi 2022) as well as chemical methods (Karnik 2021). Nevertheless, these methods have some drawbacks. The latter two methods are relatively less efficient. The former three methods, on the other hand, are slow and dangerous for the minesweeper and the animal itself. Therefore, the researchers are using sensor technology to cope with this problem and are using ground penetrating radar (GPR) (A. Marsh 2019); (Pambudi 2020) and infrared (IR) thermography (Yao 2019); (Forero-Ramirez 2022). The efficiency of these methods is promising but they are not cost-effective. Cost efficiency is a very important factor because most of the countries affected by these deadly devices are Third World countries where the dearth of funds is often a problem. So, it is essential to develop low-cost solutions to save innocent lives.

Therefore, this paper aims to propose an efficient as well as a cost-effective method for demining without posing any threat to the life of the operator through remote detection. There are three design constraints: 1) the accuracy of detection, 2) the safety of the operator, and 3) the cost-effectiveness of the system. These are achieved by using an unmanned aerial vehicle (UAV) mounted with a low-cost thermal camera to detect buried landmines through an innovative image processing algorithm. This system can also be used for surveillance purposes.

The paper is structured as follows: Section 2 offers an extensive examination of the methods already available. Section 3 elucidates the proposed solution, with subsections that delve into various components of the solution. The initial subsection presents a concise overview of the hardware, while the subsequent part provides a detailed analysis of the image processing procedures. Section 4 encompasses the experimental setup and presents the resultant findings. Finally, Section 5 consolidates the conclusions derived from the study and outlines potential future avenues for research.

## 2. Related Work:

In this section, some of the available methods in context to the design constraints are unfolded. The use of a robot with metal detection capabilities, GPR, or IR thermography seems to be a good solution to nullify the threat to the safety of the operator. These include wheeled, dragged as well as legged robots. Colorado et al. discuss some of the popular robots with their pros and cons (Colorado 2015). An advantage of these robots is their ability to handle heavier and larger equipment that typically requires substantial operational resources. Additionally, they have sufficient payload capacity to carry out these tasks effectively. However, these robotic systems face limitations when it comes to navigating uneven terrain. They may be too heavy to transport to potential minefield locations or prohibitively expensive for certain applications. These factors restrict their usability.

So, the use of a UAV is a better approach compared to the use of robots. However, this is just one part of the solution that is, ensuring the safety of the operator through remote sensing. However, the efficient detection of a landmine is a completely new challenge. There are different approaches to this part. Colorado et al. use a CMOS camera for the detection of landmine-like objects through image processing. That is, the detected object might or might not be a landmine. Moreover, it has an accuracy of nearly 80%, which is satisfactory but not much good. Šipoš et al. present the use of a ground-penetrating radar mounted on a UAV (Šipoš 2020). However, the accuracy of detection of the GPR sensor is greatly affected by the type of soil. Moreover, it cannot distinguish between a landmine and plant roots, rocks, and cuts in the land as well as animal burrows. Thus, produces false alarms in such scenarios.

So, the most promising is the use of IR thermography for the detection of the landmine. It has high accuracy and there are several image processing algorithms to elegantly accomplish this task. Now let us unfold this approach in detail.

The IR radiations lie beyond the visual light segment of the electromagnetic wave spectrum. The wavelength of IR radiations ranges between 700 nm and 1 mm. Thus, they cannot be visualised with the unaided eye. An IR sensor is used for this purpose and it detects the radiation in the form of heat. The IR sensors can be categorised into two types, that is, active IR sensors and passive IR sensors. The passive IR sensors detect the natural radiation of the object. On the other hand, the active IR sensors use a high-power source of radiation to further enhance the temperature difference of landmines (Szymanik 2016). But the landmine could explode by overheating. The basic principle behind the use of the IR sensor in landmine detection is that the thermal signature of the metallic case of the landmine is different from the surrounding soil (Nguyen 2005). So, IR thermography exploits this property and detects the landmine by detecting the change of heat signature by using an IR sensor through image processing.

Kaya et al. used thermal image time series to detect the landmines using a test minefield. The IR camera is fixed, and images are captured at different time instances (S. a. Kaya 2017). In their recent work, the authors have used a different approach and taken into account the landmines covered by plants (S. L. Kaya 2020). Unlike Kaya et al., in this work, the camera is not fixed, but the image time series is used to investigate the detection of landmines. A test minefield is also prepared to capture the real images, so that, the results can be compared.

## 3. Proposed Method:

The proposed method aims to achieve precise landmine detection using IR thermography. An IR image of the soil is captured using the FLIR Dev Kit, which is mounted on a UAV. This setup ensures the operator's safety by allowing them to remotely survey the hazardous soil from a safe distance. The FLIR Dev Kit is connected to a Raspberry Pi 3 Model B, enabling real-time image capture and transmission to the base computer. The base computer utilizes an efficient algorithm to detect landmines. An overview of the anticipated concept is depicted in Figure 1.

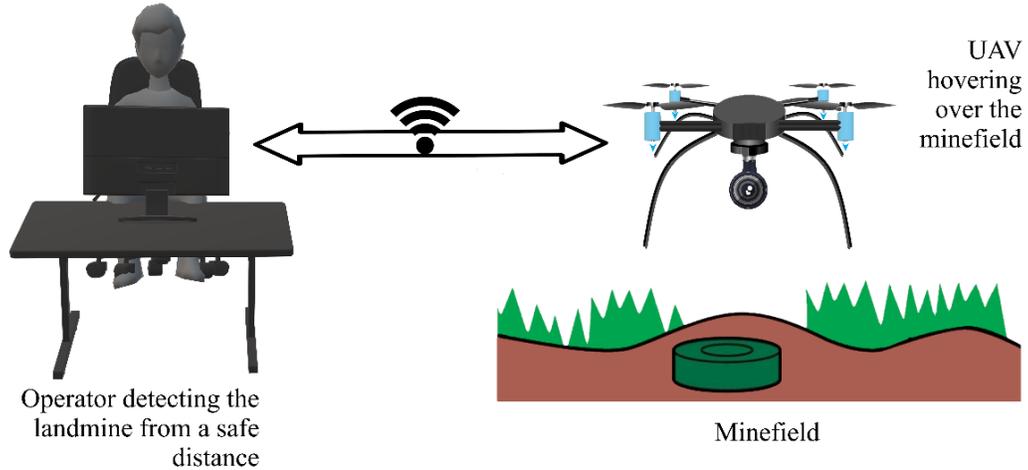

Figure 1. Synopsis of the proposed solution

### *3.1 Unmanned Aerial Vehicle (UAV)*

The presence of a stable and capable UAV is vital to ensuring operator safety within the proposed solution. It is essential for the UAV to maintain stability throughout the flight to capture clear images of the soil. Furthermore, the UAV should have the capacity to accommodate the detection module (which will be discussed in the next section) while remaining cost-effective. With these considerations in mind, a custom-built UAV has been selected as the optimal choice. The UAV is shown in Figure 2. Some specifications of the UAV are shown in Table 1.

Table 1. The UAV specifications

| Specification | Value |
|---|---|
| Payload capacity | 0.45 kg |
| Flight time | 35 – 45 minutes |
| Dimensions (LxWxH) | 25.3 x 18.3 x 5.5 cm |
| Weight | 1.21 kg |

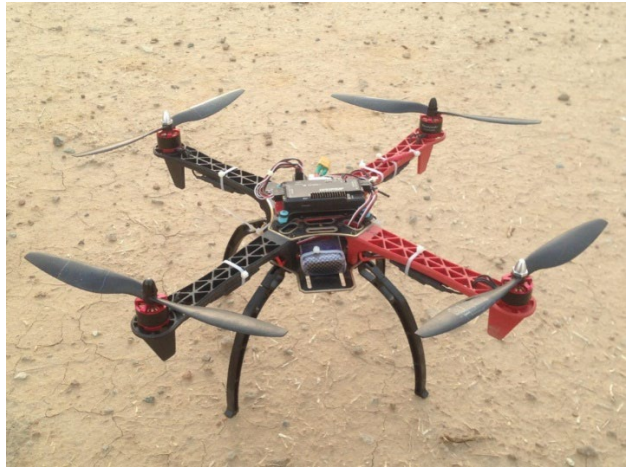

Figure 2. The custom-built UAV

*3.2 Landmine*

To assess the viability of the proposed solution, a testing landmine (TL) is purposefully constructed. The TL is cylindrical in shape, with a radius of 0.152 m and a height of 0.0635 m, aiming to replicate the structural characteristics of real landmines as accurately as possible. Although the TL does not contain any explosive materials, it is filled with wax to replicate the thermal properties typically associated with explosive substances. This choice allows for a closer approximation of the thermal behaviour exhibited by actual landmines. To simulate a testing minefield (TM), the TL is buried in the ground, as depicted in Figure 3.

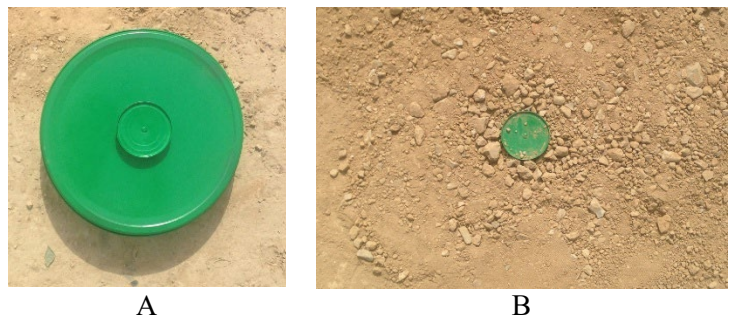

A          B
Figure 3. Testing landmine (TL): a) top view (not buried), b) buried in the soil

*3.3 Thermal Camera*

The thermal camera holds a position of utmost importance within the proposed solution. However, it also represents a significant cost factor. Opting for high-quality thermal cameras enhances the system's efficiency by providing a larger number of pixels, resulting in clear and detailed soil images. Unfortunately, this improved image quality comes at a higher price, which contradicts the cost-efficiency constraint of the solution. Furthermore, the larger pixel data necessitates more processing power and a more sophisticated computing system to execute the image processing algorithm, resulting in increased processing time. Taking into account these considerations, the FLIR Dev Kit is utilized and integrated with the Raspberry Pi 3 Model B, which operates on the Raspbian OS. These components, along with the battery, are combined to form the detection module, shown in Figure 4.

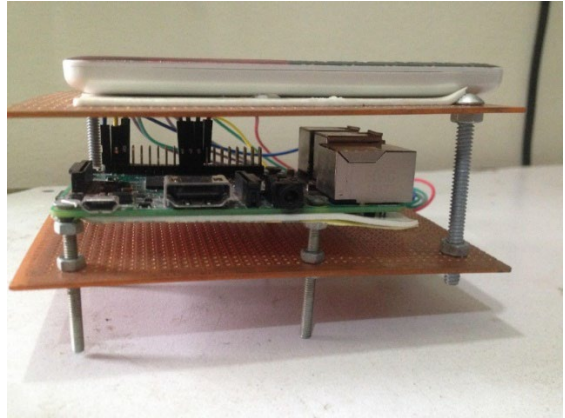

Figure 4: The Detection Module comprising the Raspberry Pi and thermal camera

This module is securely mounted beneath the UAV to capture IR images of the soil. To establish a remote connection between the Raspberry Pi and our base computer, we employ PuTTY, leveraging the SSH (Secure Shell) protocol. This configuration enables seamless remote control of the Raspberry Pi through command-line access. To achieve real-time transfer of captured images from the Raspberry Pi to the base computer, we utilize FileZilla, which ensures secure image transmission via the FTP (File Transfer Protocol). It is important to note that both the Raspberry Pi and the base computer must be connected to the same network for seamless communication.

*3.4 Image Processing of IR Image*

Once the Raspberry Pi transfers the IR image to the base computer, it undergoes image processing using a dedicated algorithm designed to identify the presence of a landmine. This algorithm comprises several distinct steps, which will be discussed in subsequent sections. A visual representation of the sequential steps can be seen in Figure 5, depicted as a flowchart illustrating the proposed algorithm.

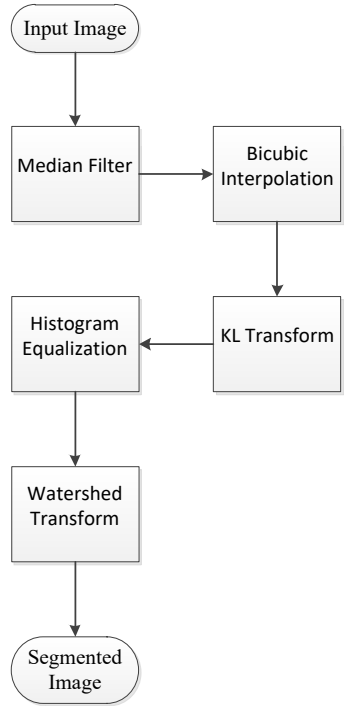

Figure 5. Flowchart of the proposed image processing algorithm

### *3.4.1 Image Filtering:*

Image noise, resulting from camera hardware and image transmission, degrades image quality. Resultantly, it adversely affects subsequent image processing, leading to increased false alarms. To address this, noise removal is crucial. The image noise can be represented as:

$$s = u + n \qquad (1)$$

Here, $s$ is the captured image with noise, $u$ is the noise-free original image, and $n$ is the additive noise. The proposed solution utilizes a median filter for noise removal. The median filter is non-linear, effectively reduces noise, preserves edges, and has low computational complexity. It operates on a pixel-by-pixel basis, replacing each pixel value with the median value of neighbouring pixels within a 3x3 window. This process is performed for each pixel, resulting in a filtered image with reduced noise. Figure 6 illustrates this process for the entry $s_{22}$ shown in red.

| 1 | 3 | 0 | 43 | 226 | 12 | 209 | ... | 181 |
|---|---|---|---|---|---|---|---|---|
| 3 | 4 | 1 | 90 | 223 | 13 | 215 | ... | 12 |
| 2 | 0 | 1 | 223 | 168 | 223 | 31 | ... | 212 |
| 0 | 2 | 3 | 113 | 208 | 82 | 123 | ... | 207 |
| 1 | 0 | 4 | 12 | 59 | 17 | 13 | ... | 223 |
| 0 | 2 | 3 | 1 | 232 | 35 | 0 | ... | 13 |
| 2 | 1 | 1 | 20 | 19 | 223 | 31 | ... | 0 |
| ⋮ | ⋮ | ⋮ | ⋮ | ⋮ | ⋮ | ⋮ |  | ⋮ |
| 2 | 1 | 0 | 33 | 119 | 248 | 113 | ... | 10 |

(a)

| 1 | 3 | 0 | 43 | 226 | 12 | 209 | ... | 181 |
|---|---|---|---|---|---|---|---|---|
| 3 | 1 | 1 | 90 | 223 | 13 | 215 | ... | 12 |
| 2 | 0 | 1 | 223 | 168 | 223 | 31 | ... | 212 |
| 0 | 2 | 3 | 113 | 208 | 82 | 123 | ... | 207 |
| 1 | 0 | 4 | 12 | 59 | 17 | 13 | ... | 223 |
| 0 | 2 | 3 | 1 | 232 | 35 | 0 | ... | 13 |
| 2 | 1 | 1 | 20 | 19 | 223 | 31 | ... | 0 |
| ⋮ | ⋮ | ⋮ | ⋮ | ⋮ | ⋮ | ⋮ |  | ⋮ |
| 2 | 1 | 0 | 33 | 119 | 248 | 113 | ... | 10 |

(b)

Figure 6. Image matrix $s$ of order $m$ by $n$: a) 3 by 3 window is selected for $s_{22}$ and is highlighted as blue, b) median value of ordered-sequence of window entries is replaced with the original value, and the new value is highlighted as blue.

It can be modelled as:

$$\hat{u}[m_x, n_y] = median\ s[i \times j]\ |\ i, j \in \breve{w} \qquad (2)$$

Whereas, $\hat{u}[m_x, n_y]$ is the pixel value in the modified image in the $m_x$ row and $n_y$ column. $\breve{w}$ is the window size specified by the $i$ number of rows and $j$ number of columns such that $\hat{u}[m_x, n_y]$ becomes the centre of the window matrix. It is important to select an odd-ordered window size for this purpose.

The same procedure is applied to all pixel values in the image. For boundary values, adjustments are made by padding rows or columns with zeros or extending the respective boundary row or column. This guarantees a consistent noise removal process across the entire image, resulting in a noise-free image represented as $\hat{u}$:

$$\hat{u} \approx s - n \qquad (3)$$

### 3.4.2 Image Resizing:

Typically, conventional IR thermography techniques do not involve image resizing since the IR cameras used provide sufficient data. However, the proposed solution utilizes a cost-effective camera to optimize system costs. The camera kit captures an 80 by 60 IR image. To enable further processing, interpolation is employed to increase the size of the IR image, resulting in an enhanced image denoted as $\hat{U}$ and is sufficiently large for further processing.

To demonstrate the resizing process, consider a 3 by 3 portion of the filtered image, with pixel values represented by alphabets. The objective is to resize this image to a 5 by 5 enhanced image. Figure 7 illustrates the original image and the initial enhanced image.

| A | B | C |
|---|---|---|
| E | F | G |
| H | I | J |

a

| A |   | B |   | C |
|---|---|---|---|---|
|   | X |   |   |   |
| E |   | F |   | G |
|   |   |   |   |   |
| H |   | I |   | J |

b

Figure 7. Image matrix enhancement: a) Original 3 by 3 portion of the image b) Enhanced image with missing entries

To fill the empty positions in the enhanced image, an interpolation operation is employed. We propose bicubic interpolation due to its ability to generate smoother interpolated images with reduced computational complexity.

Let us consider a pixel position X in the enhanced image that requires interpolation using bicubic interpolation. This involves determining the missing values among the points A, B, E, and F by fitting a unit square at this position using a third-order polynomial function. The pixel values of the corner points and the derivative values along the horizontal direction $f_x$, vertical direction $f_y$, and diagonal direction $f_{xy}$ of the unit square are necessary. The interpolated surface can be written as follows:

$$f(\hat{x}, \hat{y}) = \sum_{i=0}^{3} \sum_{j=0}^{3} a_{i,j} \, x^i \, y^j \tag{4}$$

Where $\hat{x}$ and $\hat{y}$ are the coordinates of the pixel in the transformed image, and $x$ and $y$ are the coordinates of the pixel in the original image. The interpolated surface can be obtained by finding all 16 values of coefficients $a_{i,j}$. Four coefficients are directly determined by the pixel values at the corners of the unit square. Four are calculated by the spatial derivative $f_x$ in the horizontal direction and four more are calculated by the spatial derivative $f_y$ in the vertical direction. Lastly, the remaining four coefficients are calculated by the diagonal derivative $f_{xy}$. We can write it in matrix form as follows:

$$\begin{bmatrix} a_{00} & a_{01} & a_{02} & a_{03} \\ a_{10} & a_{11} & a_{12} & a_{13} \\ a_{20} & a_{21} & a_{22} & a_{23} \\ a_{30} & a_{31} & a_{32} & a_{33} \end{bmatrix} = \begin{bmatrix} 1 & 0 & 0 & 0 \\ 0 & 0 & 1 & 0 \\ -3 & 3 & -2 & -1 \\ 2 & -2 & 1 & 1 \end{bmatrix} \begin{bmatrix} f(0,0) & f(0,1) & f_y(0,0) & f_y(0,1) \\ f(1,0) & f(1,1) & f_y(1,0) & f_y(1,1) \\ f_x(0,0) & f_x(0,1) & f_{xy}(0,0) & f_{xy}(0,1) \\ f_x(1,0) & f_x(1,1) & f_{xy}(1,0) & f_{xy}(1,1) \end{bmatrix} \begin{bmatrix} 1 & 0 & -3 & 2 \\ 0 & 0 & 3 & -2 \\ 0 & 1 & -2 & 1 \\ 0 & 0 & -1 & 1 \end{bmatrix} \tag{5}$$

Once these 16 coefficients are determined, the interpolated surface can be obtained by applying the following relation. This allows us to fill in the missing points in the enhanced image.

$$f(\hat{x}, \hat{y}) = \begin{bmatrix} 1 & x & x^2 & x^3 \end{bmatrix} \begin{bmatrix} a_{00} & a_{01} & a_{02} & a_{03} \\ a_{10} & a_{11} & a_{12} & a_{13} \\ a_{20} & a_{21} & a_{22} & a_{23} \\ a_{30} & a_{31} & a_{32} & a_{33} \end{bmatrix} \begin{bmatrix} 1 \\ y \\ y^2 \\ y^3 \end{bmatrix} \quad (6)$$

Likewise, the missing pixel values in the enhanced image can be interpolated by constructing additional bicubic surfaces at various points and seamlessly patching them together. This patching ensures that the derivative values along the adjacent boundaries are consistent and smooth.

### 3.4.3 *Feature Extraction:*

Feature extraction plays a crucial role in image processing. The Karhunen-Loève Transform (KLT), also known as Hotelling or the Principal Components Transform, is widely used for this purpose. It begins by creating a covariance matrix from the input image sequence and then extracts features based on the eigenvalues of this matrix. The transformed image, referred to as the transformed image $\hat{U}$, can be represented in a general form (Figure 8).

| $U_{11}$ | $U_{12}$ | $U_{13}$ | $U_{14}$ | $U_{15}$ | $U_{16}$ | $U_{17}$ | … | $U_{1n}$ |
|---|---|---|---|---|---|---|---|---|
| $U_{21}$ | $U_{22}$ | $U_{23}$ | $U_{24}$ | $U_{25}$ | $U_{26}$ | $U_{27}$ | … | $U_{2n}$ |
| $U_{31}$ | $U_{32}$ | $U_{33}$ | $U_{34}$ | $U_{35}$ | $U_{36}$ | $U_{37}$ | … | $U_{3n}$ |
| $U_{41}$ | $U_{42}$ | $U_{43}$ | $U_{44}$ | $U_{45}$ | $U_{46}$ | $U_{47}$ | … | $U_{4n}$ |
| $U_{51}$ | $U_{52}$ | $U_{53}$ | $U_{54}$ | $U_{55}$ | $U_{56}$ | $U_{57}$ | … | $U_{5n}$ |
| $U_{61}$ | $U_{62}$ | $U_{63}$ | $U_{64}$ | $U_{65}$ | $U_{66}$ | $U_{67}$ | … | $U_{6n}$ |
| $U_{71}$ | $U_{72}$ | $U_{73}$ | $U_{74}$ | $U_{75}$ | $U_{76}$ | $U_{77}$ | … | $U_{7n}$ |
| ⋮ | ⋮ | ⋮ | ⋮ | ⋮ | ⋮ | ⋮ | | ⋮ |
| $U_{m1}$ | $U_{m2}$ | $U_{m3}$ | $U_{m4}$ | $U_{m5}$ | $U_{m6}$ | $U_{m7}$ | … | $U_{mn}$ |

Figure 8. General representation of the enhanced image $\hat{U}$

To apply the KLT, the data must be represented in terms of the collection of vectors. That is why; every column of this matrix is split into *n* number of column vectors *y*, that is:

$$y_1 = \begin{bmatrix} U_{11} \\ U_{21} \\ U_{31} \\ U_{41} \\ U_{51} \\ U_{61} \\ U_{71} \\ \vdots \\ U_{m1} \end{bmatrix}, \; y_2 = \begin{bmatrix} U_{12} \\ U_{22} \\ U_{32} \\ U_{42} \\ U_{52} \\ U_{62} \\ U_{72} \\ \vdots \\ U_{m2} \end{bmatrix}, \; \cdots, \; y_n = \begin{bmatrix} U_{1n} \\ U_{2n} \\ U_{3n} \\ U_{4n} \\ U_{5n} \\ U_{6n} \\ U_{7n} \\ \vdots \\ U_{mn} \end{bmatrix} \quad (7)$$

Then the mean vector of order $m$ by 1 and covariance matrix of order $m$ by $n$ is calculated as follows:

$$\mu_x = \frac{1}{n} \sum_{i=0}^{n} y_i \quad (8)$$

$$c_x = \frac{1}{n} \sum_{i=0}^{n} (y_i - \mu_i)(y_i - \mu_i)^T \quad (9)$$

For the covariance matrix $c_x$, we find eigenvalues $\lambda_i$ where $i = 0, 1, 2, \ldots, n-1$. These eigenvalues are ordered in descending order that is $\lambda_0 \geq \lambda_1 \geq \lambda_2 \geq \cdots \geq \lambda_{n-1}$. Corresponding to every eigenvalue we find an eigenvector $e_i$ where $i = 0, 1, 2, \ldots, n-1$. Then the final transformation matrix is obtained as:

$$A = \begin{bmatrix} e_0^T \\ e_1^T \\ e_2^T \\ e_3^T \\ e_4^T \\ \vdots \\ e_{n-1}^T \end{bmatrix} \quad (10)$$

We use this transformation matrix A to get the transformed image as follows:

$$K = A(\hat{U} - \mu_x) \quad (11)$$

This transformation establishes a new coordinate system with its origin at the center of the landmine. The axes of this coordinate system align with the direction of the eigenvectors. This alignment along the eigenvectors makes the detection process easier as the data is well-aligned and oriented.

### 3.4.4 Contrast Enhancement:

In order to improve the visibility of the landmine in the transformed image obtained from KLT, contrast enhancement is performed. Histogram Equalization is used for this purpose due to its high efficiency. The histogram of an image provides a global description of its intensity distribution. The digital image has different $L$ intensity values, 0 to $L-1$. If, in the transformed image $K$, $r_k$ represents the $k_{th}$ intensity level, then the histogram of the image can be defined as follows:

$$p(r_k) = \frac{\text{Number of pixels having intensity value } r_k}{\text{Total number of pixels in the image}} \tag{12}$$

$$p(r_k) = \frac{n_k}{n} \; ; \quad 0 \leq k \leq L-1 \tag{13}$$

This histogram is normalized to $[0, 1]$. Since the image is digitized using 8 bits, the value of the intensity level $L$ is 256. The $p(r_k)$ actually tells us the probability of occurrence of a pixel having the intensity value equal to $r_k$. Generally, the histogram of a low-contrast image has a very small span of intensity values. The histogram equalization transform $T(r_k)$ tends to increase the dynamic range of intensity values by taking the cumulative distribution function (CDF) of $r$ and thus produces a transformed high-contrast image as follows:

$$f_k = T(r_k) = \sum_{i=0}^{k} p_r(r_i) \tag{14}$$

Where $r_k$ is the intensity of $k_{th}$ pixel in the original image and $f_k$ is the intensity of the same pixel in the transformed image $F$. The above equation can be written as:

$$f_k = T(r_k) = \sum_{i=0}^{k} \frac{n_i}{n} \tag{15}$$

The transform $T(r_k)$ possesses the following two properties:
  I. $T(r_k)$ is single-valued as well as monotonically increases in the range $0 \leq r \leq L-1$.
  II. The range of $T(r_k)$ is $0 \leq T(r_k) \leq L-1$ for the domain $0 \leq r \leq L-1$.

The first condition maintains the order of grey levels in the transformed image and the second condition ensures that the transformed image does not possess any intensity value beyond the scope of the image.

*3.4.5 Segmentation:*

Segmentation is a crucial step in image analysis, as it allows the image to be divided into meaningful parts or objects. It involves transforming the image into segments or regions based on the similarities between pixel information.

There are two main approaches to segmentation: the boundary-based approach and the region-based approach. The boundary-based approach detects explicit or implicit boundaries between different regions by identifying local changes in the image. In contrast, the region-based approach segments the image into different regions based on similarities in pixel values or other criteria. In our proposed solution, we employ the watershed transform for image segmentation, which falls under the region-based approach.

The watershed transform treats the image as a natural geological watershed, where the brighter grey levels represent higher altitudes and the darker grey levels represent lower levels in the topographic area. In this transformation, the grey-tone image is represented by a function $f(x)$ that assigns a grey level value to each point $x$ in the image. It is defined as $f(x): Z^2 \to Z$.

Now let us define an important parameter namely the minima. Let us consider two distinct points $s_1(x_1, f(x_1))$ and $s_2(x_2, f(x_2))$ on this topographic surface $S$. Corresponding to these two points, we can define a non-ascending path consisting of a sequence $\{s_i\}$ as follows:

$$\forall s_i(x_i, f(x_i)), s_j(x_j, f(x_j)) \quad i \geq j \Leftrightarrow f(x_i) \leq f(x_j) \tag{16}$$

A point is called minima if and only if there does not exist any non-ascending path between a beginning point $s_b(x_b, f(x_b))$ and an ending point $s_e(x_e, f(x_e))$ such that $f(x_e) < f(x_b)$. As there might be more than one valley in our topographical image, so, there will be more than one local minima which collectively can be represented through a set $m(f)$. Next, we flood these different valleys from their respective minima. As water from different valleys increases, they tend to merge and the number of valleys decreases. However, we prevent this merging by using dams as barriers. The flooding process stops when the highest peak is submerged. The result is a landscape consisting of different catchment basins represented as $CB_i(f)$, each represented by a unique minima $m_i(f)$. These catchment basins are separated by watersheds (dams), yielding the desired segmented image.

## 4. Experimental Results:

The proposed method was tested on the TM. The image processing algorithm was executed using MATLAB 2017. Figure 9 shows a snap of the testing process in which the detection module is integrated with the UAV and the UAV is capturing the IR images of the soil.

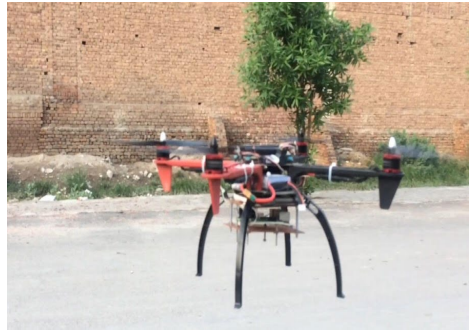

Figure 9. UAV capturing images during testing.

The image time series was formed by capturing four pictures at a time from 6 AM to 10 AM and 5 PM to 10 PM after every hour. Thus, 44 images were captured for a single location. The images were not captured from 11 AM to 4 PM, because the camera could not capture the IR image of the soil, instead, it captured the IR image of the sun rays. The Raspberry pi worked as a local server and communicated with the base computer through a static IP. Thus, the captured images were transferred to the base computer in real time.

The received images at the base computer were filtered out using the median filter. Though not all the images were affected by the noise, this step must be performed as a precaution. Since the images had a low resolution, their size was enhanced through bicubic interpolation. We found the optimal resolution to be 184 by 168. The images were not enhanced further, because this would affect the symmetry of the image, and the implicit boundary between the landmine and the soil might fade away. The large number of images in the thermal image time series was reduced to one through KLT using information gathered from all the images. Figure 10(a) shows the image obtained after KLT. The resultant image was subjected to contrast enhancement through histogram equalization to enhance the difference between the landmine and the surrounding soil, as shown in Figure 10(b). Finally, the watershed transform was applied for the segmentation to evidently inspect the landmine in the soil, as shown in Figure 10(c). The performance of the proposed solution was tested on the following parameters:

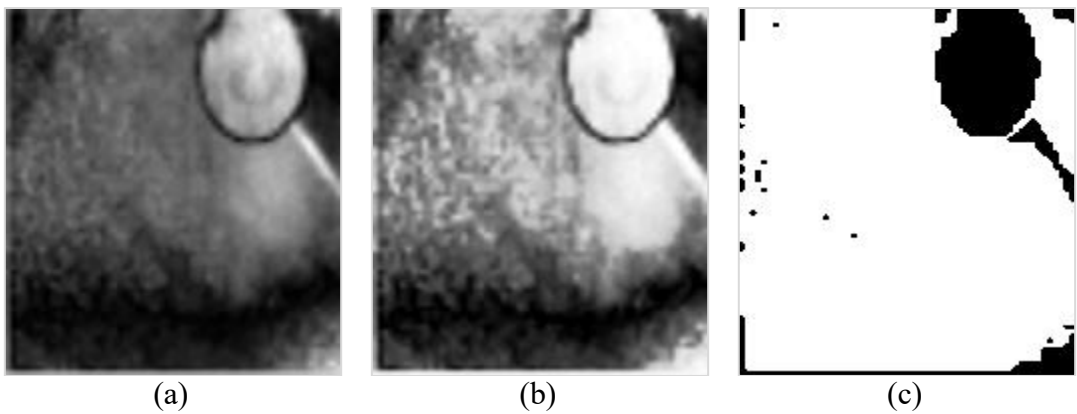

(a)          (b)          (c)

Figure 10. Image processing results of (a) KLT, (b) histogram equalization, and (c) watershed transform.

    I.    True positive (TP): There is a landmine in the soil and it was detected correctly.

    II.    False positive (FP): There is not a landmine in the soil but it was erroneously detected.

    III.    True negative (TN): There is not a landmine in the soil and it did not detect either.

    IV.    False negative (FN): There is a landmine in the soil but it is missed.

There were 50 trials performed for each of the former two parameters and 44 of them were TP, that is, 88% of landmines were detected correctly and only 12% were missed (FN). The algorithm was subjected to the images with no landmines and 100% were TN and none were FP. Figure 11 shows the segmented image of an FN trial. There is not any symmetry in the segments which shows the absence of a landmine. The vague segments illustrate the dust pattern.

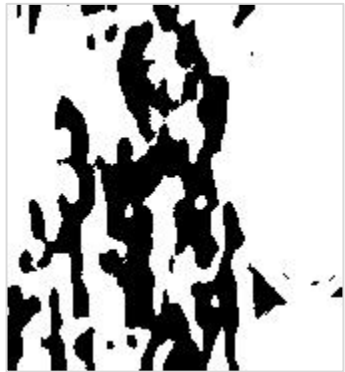

Figure 11. The resultant image of an FN trial.

The altitude at which the UAV operates directly influences the performance of the system. When the UAV flies too close to the ground, it can disturb loose small-grain soil, resulting in both positive and negative impacts on the detection process. On the positive side, this disturbance can clear the soil covering a buried landmine, making it more visible and potentially increasing the chances of detection. However, this process also has drawbacks that affect the image-capturing process. At such a low altitude, the camera captures the heat signature of the dirt particles floating in the air instead of the heat signature of the soil. This can have adverse effects on image processing and analysis.

Furthermore, the floating dirt particles can pose a risk to the UAV itself. These particles can stick to the UAV's motors, causing friction and potentially damaging them. If left uncleaned, the motors may overheat due to increased friction. It is important to note that these concerns primarily apply to loose small-grain soil. In contrast, if the soil is more compact or has moisture, the issue of floating dirt particles is less prominent and may not pose a problem.

On the other hand, when the altitude was too high, the camera could not capture an accurate picture of the soil. The system was tested on different altitudes, and Figure 12 shows the resultant segmented images of a TP trial at 1m, 1.5m, and 2m altitudes. There were 10 trials conducted for these altitudes, Figure 13 shows the resultant plot of the rate of detection for different altitudes. It signifies that the altitude at which UAV flies directly affects the detection process and the optimal altitude would depend upon the camera resolution. In our case, we obtained optimal results at 1.5m altitude.

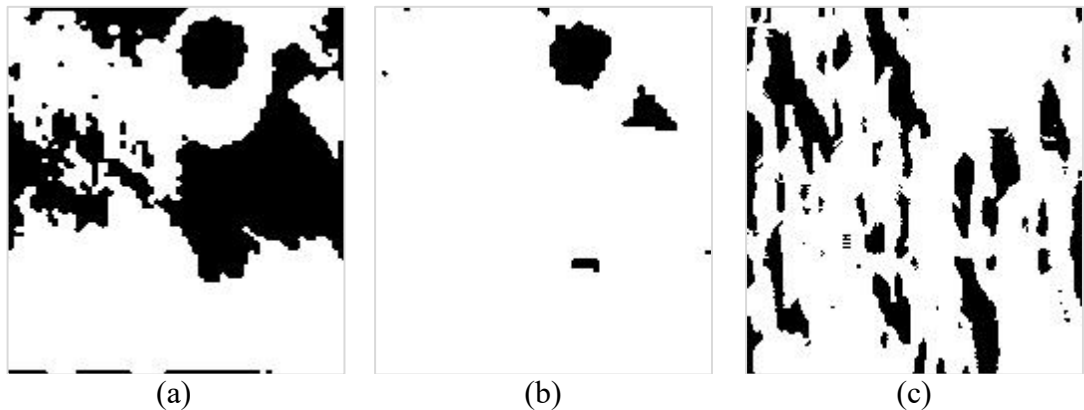

(a) (b) (c)
Figure 12. The resultant image of a TP trial for an altitude of (a) 1m, (b) 1.5m, and (c) 2m.

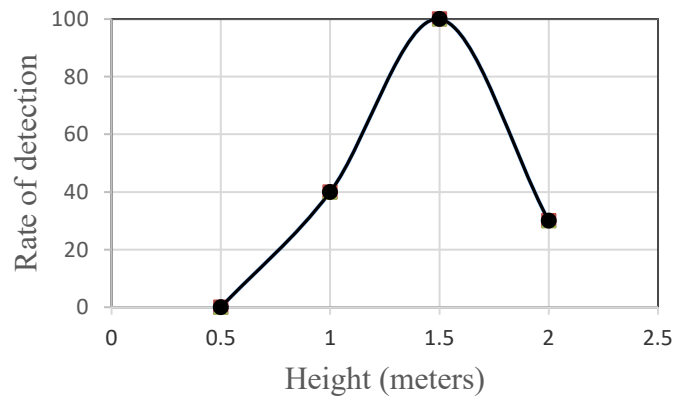

Figure 13. The graph of detection rate for different altitudes of the UAV.

Kaya et al. used FLIR T 650 SC camera and ATOMTM 1024 camera. Both of these are very expensive. Moreover, these cameras were fixed at a certain altitude rather than integrating with a UAV. Thus it did not consider the effect of altitude at which images were captured. They used a relatively complex method for image processing and it requires more processing and a sophisticated machine. They have claimed to achieve 91% and 95% accuracies in the case of raw and filtered data respectively.

This work, on the other hand, used a relatively less complex algorithm to detect the landmines. There is some diversity of practical nature as well. We used a very cost-effective camera. The camera was integrated with a UAV which could cover a larger area and is a more practical approach. This study also investigated the effect of the camera's altitude on landmine detection and took into consideration the camera's stability as well. The effect of the time at which the detection process is carried out was also discussed. Overall, this research makes substantial progress in the field of landmine detection by improving upon existing methodologies and addressing practical challenges. The achieved accuracy rate of 88% demonstrates the competitiveness of the proposed solution, both in theoretical advancements and practical applicability. By embracing more realistic scenarios and minimizing reliance on assumptions, this study paves the way for more efficient and effective landmine detection systems.

## 5. Conclusion and future work:

Landmines are a curse to humanity. These deadly devices not only cost tens of thousands of innocent lives but also put a huge burden on the economy of that country. This paper introduced a UAV-based approach that addresses this menace, prioritizing accuracy, operator safety, and cost-effectiveness. By integrating an infrared (IR) camera module, the UAV captures images of the minefield, which are then processed using an advanced algorithm to detect landmines. The proposed solution underwent testing on a test minefield, demonstrating an impressive detection accuracy exceeding 88%, offering great promise. It is anticipated that the implementation of this method will contribute significantly to the preservation of hundreds of innocent lives.

Through the combination of UAV technology, image processing, and careful algorithm design, this research aims to combat the menace of landmines and mitigate their dire consequences. By continuing to refine and expand upon these efforts, we can work towards creating a safer and more secure future for all.

In addition to the aforementioned findings, it is important to note that the proposed method specifically focuses on the detection of metallic landmines. Given the increasing use of plastic landmines, it is recommended to explore detection techniques for these types of devices as well. Further investigation into plastic landmine detection would enhance the applicability and effectiveness of the proposed approach.

Furthermore, to improve operational efficiency, it is suggested that the UAV flying operation be automated. This automation would allow users to define the minefield area solely using Google Maps, enabling the UAV to autonomously sweep the designated region for landmine detection.

By streamlining the process through automation, the proposed method can become more user-friendly and efficient in addressing the challenge of landmine detection.

## 6. References

A. Marsh, L., Van Verre, W., L. Davidson, J., Gao, X., JW Podd, F., J. Daniels, D. and J. Peyton, A. 2019. "Combining electromagnetic spectroscopy and ground-penetrating radar for the detection of anti-personnel landmines." *Sensors* 19 (15): 3390.

Adee Schoon, Michael Heiman, Håvard Bach, Terje Groth Berntsen, Cynthia D. Fast. 2022. "Validation of technical survey dogs in Cambodian mine fields." *Applied Animal Behaviour Science* 251: 105638.

Benjamin Shemer, Etai Shpigel, Carina Hazan, Yossef Kabessa, Aharon J. Agranat, Shimshon Belkin. 2021. "Detection of buried explosives with immobilized bacterial bioreporters." *Microbial biotechnology* 14 (1): 251-261.

n.d. "Children and Landmines: A Deadly Legacy." *www.unicef.org.* Accessed 11 09, 2019. https://www.unicef.org/french/protection/files/Landmines_Factsheet_04_LTR_HD.pdf.

Colorado, J., Mondragon, I., Rodriguez, J. and Castiblanco, C. 2015. "Geo-mapping and visual stitching to support landmine detection using a low-cost UAV." *International Journal of Advanced Robotic Systems* 12 (9): 125.

DeAngelo, Darcie. 2018. "Demilitarizing disarmament with mine detection rats." *Culture and Organization* 24 (4): 285-302.

F.Y.C. Albert, C.H.S. Mason, C.K.J. Kiing, K.S. Ee, K.W. Chan. 2014. "Remotely operated solar-powered mobile metal detector robot." Procedia computer science.

2003. *Facts About Land Mines.* 10 16. Accessed 11 9, 2019. https://www.care.org/emergencies/facts-about-land-mines.

2018. *Facts About Landmines.* 01 15. Accessed 11 09, 2019. https://landminefree.org/facts-about-landmines/.

Facts About Landmines. 2018. "landminefree.org." 01 15. Accessed 06 11, 2023. https://landminefree.org/facts-about-landmines/.

Forero-Ramirez, J.C., García, B., Tenorio-Tamayo, H.A., Restrepo-Girón, A.D., Loaiza-Correa, H., Nope-Rodríguez, S.E., Barandica-Lopez, A. and Buitrago-Molina, J.T. 2022. "Detection of "legbreaker" antipersonnel landmines by analysis of aerial thermographic images of the soil." *Infrared Physics & Technology* 125: 104307.

Janja Filipi, Vladan Stojnić, Mario Muštra, Ross N. Gillanders, Vedran Jovanović, Slavica Gajić, Graham A. Turnbull, Zdenka Babić, Nikola Kezić, Vladimir Risojević. 2022. "Honeybee-based biohybrid system for landmine detection." *Science of the Total Environment* 803: 150041. doi:https://doi.org/10.1016/j.scitotenv.2021.150041.

Jian, L., Yang, X., Zhou, Z., Zhou, K. and Liu, K. 2018. "Multi-scale image fusion through rolling guidance filter." *Future Generation Computer Systems* 83: 310-325.

Karnik, S. and Prabhu, R. 2021. "Chemical detection of explosives in soil for locating buried landmine." Counterterrorism, Crime Fighting, Forensics, and Surveillance Technologies.

Kaya, S. and Leloglu, U.M. 2017. "Buried and surface mine detection from thermal image time series. ." *IEEE Journal of Selected Topics in Applied Earth Observations and Remote Sensing* 10 (10): 4544-4552.

Kaya, S., Leloglu, U.M. and Tumuklu Ozyer, G. 2020. "Robust landmine detection from thermal image time series using Hough transform and rotationally invariant features." *International Journal of Remote Sensing* 41 (2): 725-739.

Landmine Facts. 2016. "www.landminesurvivors.org." 07 11. Accessed 06 11, 2023. http://landminesurvivors.org/what_landmines.html.

Landmines Monitor 2022. 2022. "Landmine and Cluster Munition Monitor." 17 11. Accessed 06 11, 2023. http://www.the-monitor.org/media/3352351/2022_Landmine_Monitor_web.pdf.

Manley, Paul V. 2016. *Plant functional trait and hyperspectral reflectance responses to Comp B exposure: efficacy of plants as landmine detectors.* Virginia Commonwealth University.

Nguyen, T.T., Hao, D.N., Lopez, P., Cremer, F. and Sahli, H. 2005. "Thermal infrared identification of buried landmine." Detection and Remediation Technologies for Mines and Minelike Targets X, SPIE.

Pambudi, A.D., Fauß, M., Ahmad, F. and Zoubir, A.M. 2020. ".Minimax robust landmine detection using forward-looking ground-penetrating radar." *IEEE Transactions on Geoscience and Remote Sensing* 58 (7): 5032-5041.

Shubha Rani Sharma, Debasish Kar. 2019. "An Insight into plant nanobionics and its applications." *Plant Nanobionics* 1 (Advances in the Understanding of Nanomaterials Research and Applications): 65-82.

Šipoš, D. and Gleich, D. 2020. "A lightweight and low-power UAV-borne ground penetrating radar design for landmine detection." *Sensors* 20 (8): 2234.

Szymanik, B. and Woloszyn, M. 2016. "Magnetic and infrared thermography methods in detection of antipersonnel landmines." *COMPEL: The International Journal for Computation and Mathematics in Electrical and Electronic Engineering* 35 (4): 1323-1337.

Yao, Y., Wen, M. and Wang, Y. 2019. "Multi-temporal IR thermography for mine detection." 10th International Workshop on the Analysis of Multitemporal Remote Sensing Images (MultiTemp), IEEE.

Yip., Ed. K. R. Rao and P.C. 2001. http://citeseerx.ist.psu.edu/viewdoc/download?doi=10.1.1.701.4000&rep=rep1&type=pdf .